%% file: na61.tex
\documentclass{PoS}
\usepackage{upgreek}
\usepackage{url}
\usepackage{color}
\usepackage{amsmath}
\usepackage{mcite}
\usepackage{rviewport}
\usepackage{xspace}
\usepackage{wrapfig}
\usepackage{multicol}
\include{na61Abbreviations}

\definecolor{rossoCP3}{cmyk}{0,.88,.77,.40}
\definecolor{darkBlue}{rgb}{0, 0, 0.8}
\title{New Results from the Cosmic-Ray Program of the \\NA61/SHINE facility at the CERN SPS}

\ShortTitle{New Results from NA61/SHINE}

\author{\speaker{Michael Unger} for the \NASixtyOne
  Collaboration\thanks{http://shine.web.cern.ch/content/author-list}\\
  Karlsruhe Institute of Technology (KIT), Postfach 3640, D-76021
  Karlsruhe, Germany\\ E-mail: \email{michael.unger@kit.edu}}

\abstract{The NA61/SHINE experiment at the SPS accelerator at CERN is
  a unique facility for the study of hadronic interactions at fixed
  target energies. The data collected with NA61/SHINE is relevant for
  a broad range of topics in cosmic-ray physics including
  ultrahigh-energy air showers and the production of secondary
  nuclei and anti-particles in the Galaxy.

Here we present an update of the measurement of the momentum spectra
of anti-protons produced in $\pi^-$+C interactions at 158 and
350~\GeVc and discuss their relevance for the understanding of muons
in air showers initiated by ultrahigh-energy cosmic rays.

Furthermore, we report the first results from a three-day pilot run aimed
at investigating the capability of our experiment to measure nuclear
fragmentation cross sections for the understanding of the propagation
of cosmic rays in the Galaxy. We present a preliminary measurement of
the production cross section of Boron in C+p interactions at 13.5~\AGeVc
and discuss prospects for future data taking to provide the
comprehensive and accurate reaction database of nuclear fragmentation
needed in the era of high-precision measurements of Galactic cosmic
rays.}

\FullConference{36th International Cosmic Ray Conference -ICRC2019-\\
		July 24th - August 1st, 2019\\
		Madison, WI, U.S.A.}

\begin{document}
\section{Introduction}

\NASixtyOne\footnote{SHINE = SPS Heavy Ion and Neutrino
  Experiment}~\cite{Abgrall:2014fa} is a multi-purpose fixed target
experiment to study hadron production in hadron-nucleus and
nucleus-nucleus collisions at the CERN Super Proton Synchrotron (SPS).

The main part of the detector consists of a set of large-acceptance Time
Projection Chambers (TPCs) and two superconducting magnets with a
combined bending power of 9~Tm resulting in a precise measurement of
particle momenta ($\sigma(p)/p^2\approx
(0.3{-}7){\times}10^{-4}\,\GeVc^{-1}$) and providing
excellent particle identification capabilities via the specific energy
loss in the TPC volumes.

The experiment started operating in 2007 and has since collected
hadroproduction data with a variety of projectiles, beam energies and
target materials. Two data taking campaigns were dedicated
specifically to cosmic-ray research and will be discussed in more
detail in the next two sections. In addition, \NASixtyOne performed a
detailed measurement of particle production in p+C interactions at
31~\GeVc~\cite{Abgrall:2011ae, Abgrall:2011ts, Abgrall:2015hmv} and
$\pi^+$+C at 60~\GeVc~\cite{piC} to determine the
beam properties in accelerator-based neutrino experiments. Since
carbon is a good proxy for interactions on nitrogen in air, the measured
spectra in these reactions are also highly relevant for the modeling
of low energy interaction in air showers~\cite{Meurer:2005dt,
  Unger:2010ze}. Furthermore, p+p interactions were measured in a wide
range of beam momenta (20, 31, 40, 80 and
158~\GeVc)~\cite{Abgrall:2013pp_pim, Aduszkiewicz:2017sei} with the
aim of defining a reference data set for the heavy ion physics program
of \NASixtyOne. This data set is also very useful for the study of the
secondary production of
anti-protons~\cite{Korsmeier:2018gcy,Cuoco:2019kuu,
  Kachelriess:2019ifk} and anti-nuclei~\cite{vonDoetinchem:2016szz,
  Gomez-Coral:2018yuk} in collisions of cosmic-ray protons with the
interstellar medium in the Galaxy.

\section{Pion Interactions in Air Showers}

A long-standing problem in the physics of ultrahigh-energy cosmic rays
is the understanding of muons in air showers, which are found to be
more abundant in data than in simulated air showers
(see~\cite{wgicrc19} for a summary of recent measurements). The
majority of muons in air showers is created in decays of charged pions
when the energy of the pion is low enough such that its decay length
is smaller than its interaction length. At ultra-high energies it
takes several generations of interactions until the average pion
energy is sufficiently small to allow for pion decays. As a result,
the total number of muons in an air shower depends on details of
hadronic interactions along a chain of interactions. Hence even small
differences in the assumed properties of hadronic interactions can
lead to a sizable effect on the predicted muon number when escalated
over several generations in the particle cascade (see also discussion
in~\cite{cazonicrc19}).

The vast majority of hadronic interactions in air showers are
$\pi$+air interactions and the muons observed by particle detectors of
surface detectors mostly originate from decays of pions that got
produced in pion-air interactions at equivalent beam energies below a
TeV~\cite{Drescher:2002vp,Meurer:2005dt, Maris:2009uc}. Therefore, new
data with pion beams at 158 and 350\,\GeVc on a thin carbon target (as
a proxy for nitrogen) were collected by the \NASixtyOne experiment and
we previously reported on the production spectra of charged hadrons
and identified particles (\pions, \kaons, \protons, \lambs, \kzeros,
\kaonstar, $\omega$ and \rhozero)~\cite{unger11, hans, Herve:2015lra,
  Aduszkiewicz:2017anm, Prado:2017hub,Prado:2018wsv}.

The key to model muon production in air showers is to correctly
predict the fraction $f$ of the energy that remains in the hadronic
cascade in each interaction and is not lost to the electromagnetic
component via $\pi^0$ production. In a simplified model with the
production of only charged and neutral pions, this fraction is $f=2/3$
and after $n$ interactions $(2/3)^n$ of the initial energy is left in
the hadronic component. Muons are produced when the pions reach low
energies and decay, which happens at about $n=8$ interactions for an
air shower of $10^{20}$~eV~\cite{Matthews:2005sd}. In a more realistic
scenario the energy transfer to the hadronic component to $f \sim (2/3
+ \Delta)$, where $\Delta$ accounts for hadronic particles without
dominant electromagnetic decay channels such as \rhozero
mesons~\cite{Drescher:2007hc,Ostapchenko:2013pia} or
baryons~\cite{Pierog:2006qv}. Then a fraction of $(2/3 +
\Delta)^n\approx (2/3)^n\,(1 + 3/2\; n\, \Delta)$ of the initial
cosmic-ray energy can produce muons after $n$ interactions and only if
the value of $\Delta$ is accurately known throughout the whole chain of
interactions, there is hope for a precise prediction of the muon
number in air showers.

\begin{figure}[t]
  \centering
  \includegraphics[clip,rviewport=0.5 0 1 1, width=0.65\linewidth]{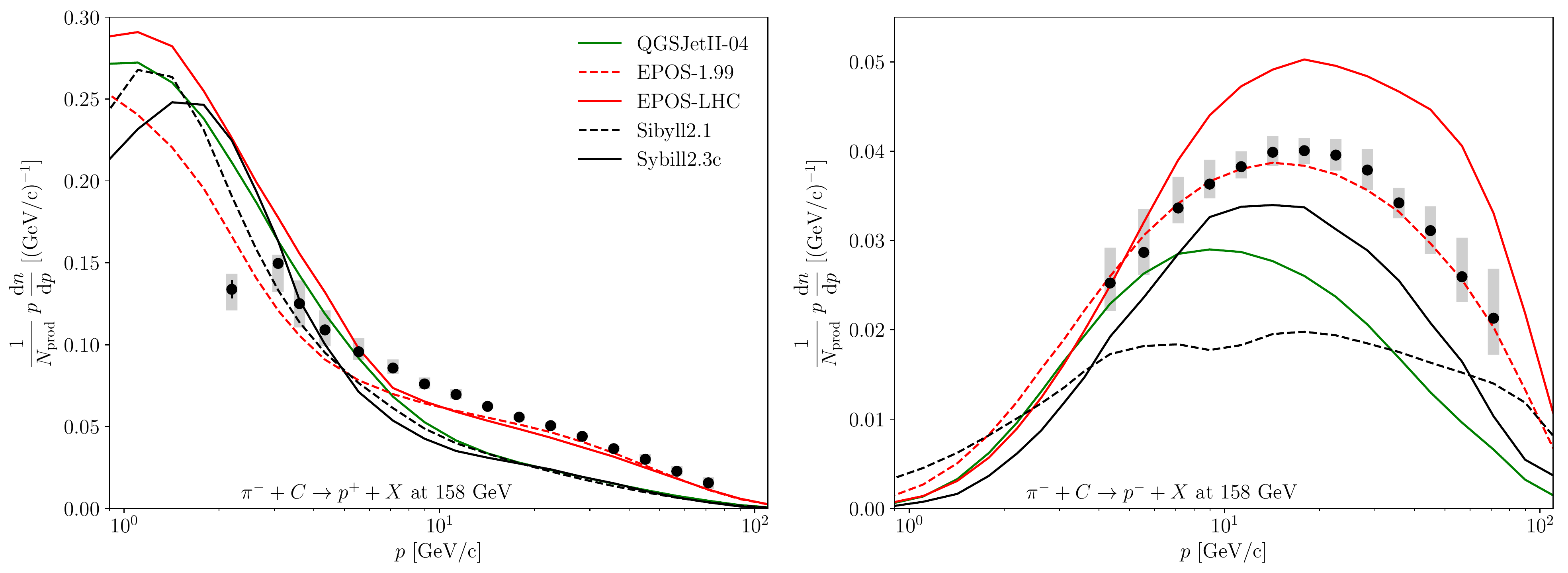}
  \includegraphics[clip,rviewport=0.3 0.3 0.48 0.96, width=0.2\linewidth]{pics/antip.pdf}
  \caption{Comparison of the \pT-integrated \antiproton spectrum at
    158~\GeVc with predictions of hadronic interaction models used in air shower simulations.}
  \label{fig:antip}
\end{figure}

A good proxy for the baryons produced in interactions (as opposed to
being struck off the target) are anti-baryons. Our previous
preliminary measurements of \antiproton production in $\pi^-+C$
interactions had sizable systematic uncertainties due to a
model-dependent correction for the feed-down of antiprotons from
electroweak decays, mainly from $\bar{\Lambda} \rightarrow
\bar{\text{p}} + \pi^-$. Here we present an update of this measurement
with a data-driven correction for the feed-down. For this purpose we
first measured the $\bar{\Lambda}$ production in $\pi^-+C$
interactions~\cite{Prado:2018wsv} and then re-weighted the model
correction accordingly as described in~\cite{Abgrall:2015hmv}. The
result is shown in Fig.~\ref{fig:antip} and compared to predictions of
hadronic interaction models used in air shower simulations. As can be
seen, none of the most recent versions of the models describes the
data well, the older version 1.99 of EPOS, however, gives a very good
description of our data. The current ``LHC'' version of this model
overproduces anti-protons and hence the enhanced transfer to the
hadronic cascade and the corresponding larger amount of muon
production in this model is strongly disfavored by our data. On the
other hand, all the remaining models under-produce anti-protons and
thus it can be conjectured that an increase of anti-baryon production
in these models to match our measurement could alleviate the muon
deficit reported by air shower experiments.

\begin{figure}[t]
 \includegraphics[width=0.5\linewidth]{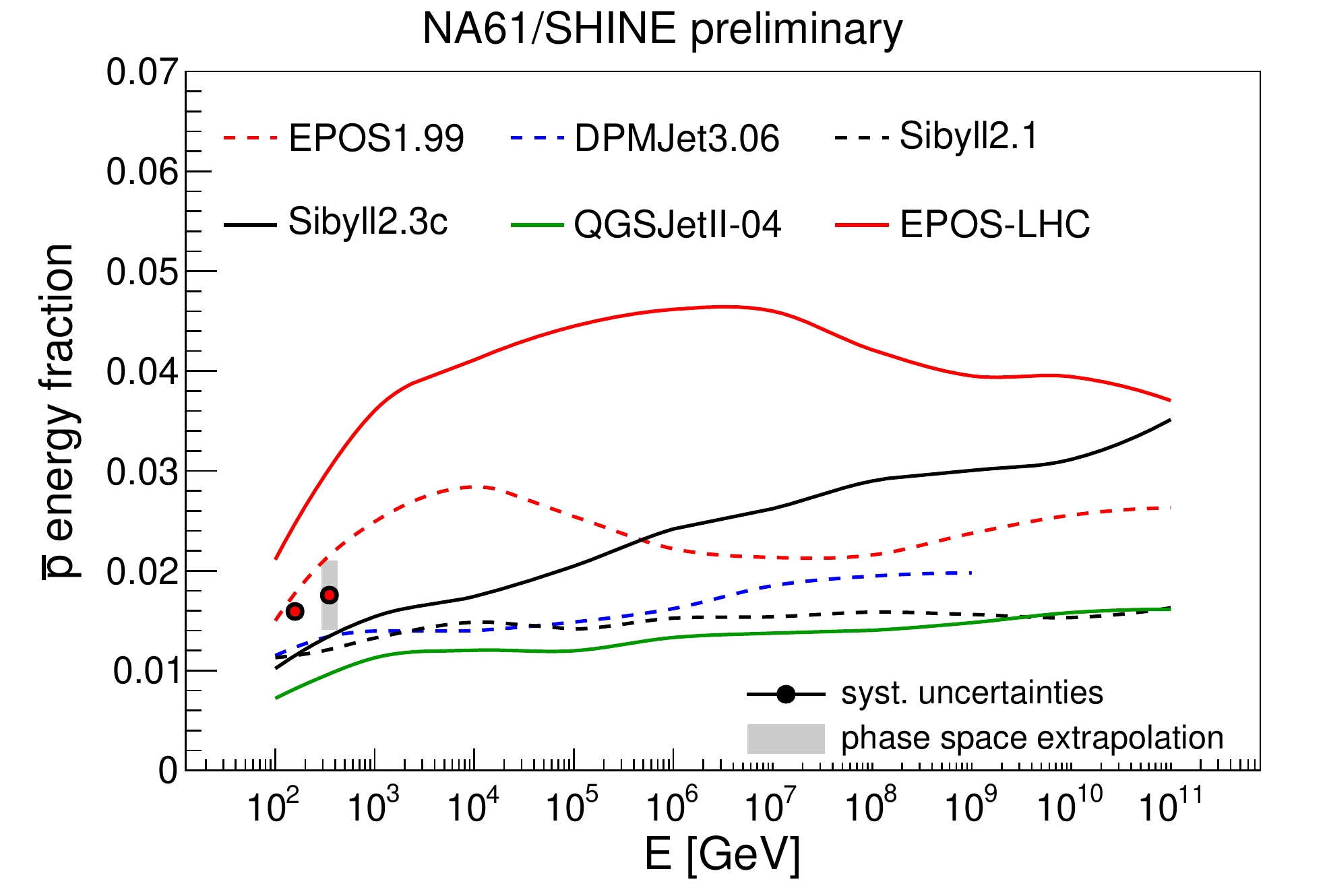}
 \includegraphics[width=0.5\linewidth]{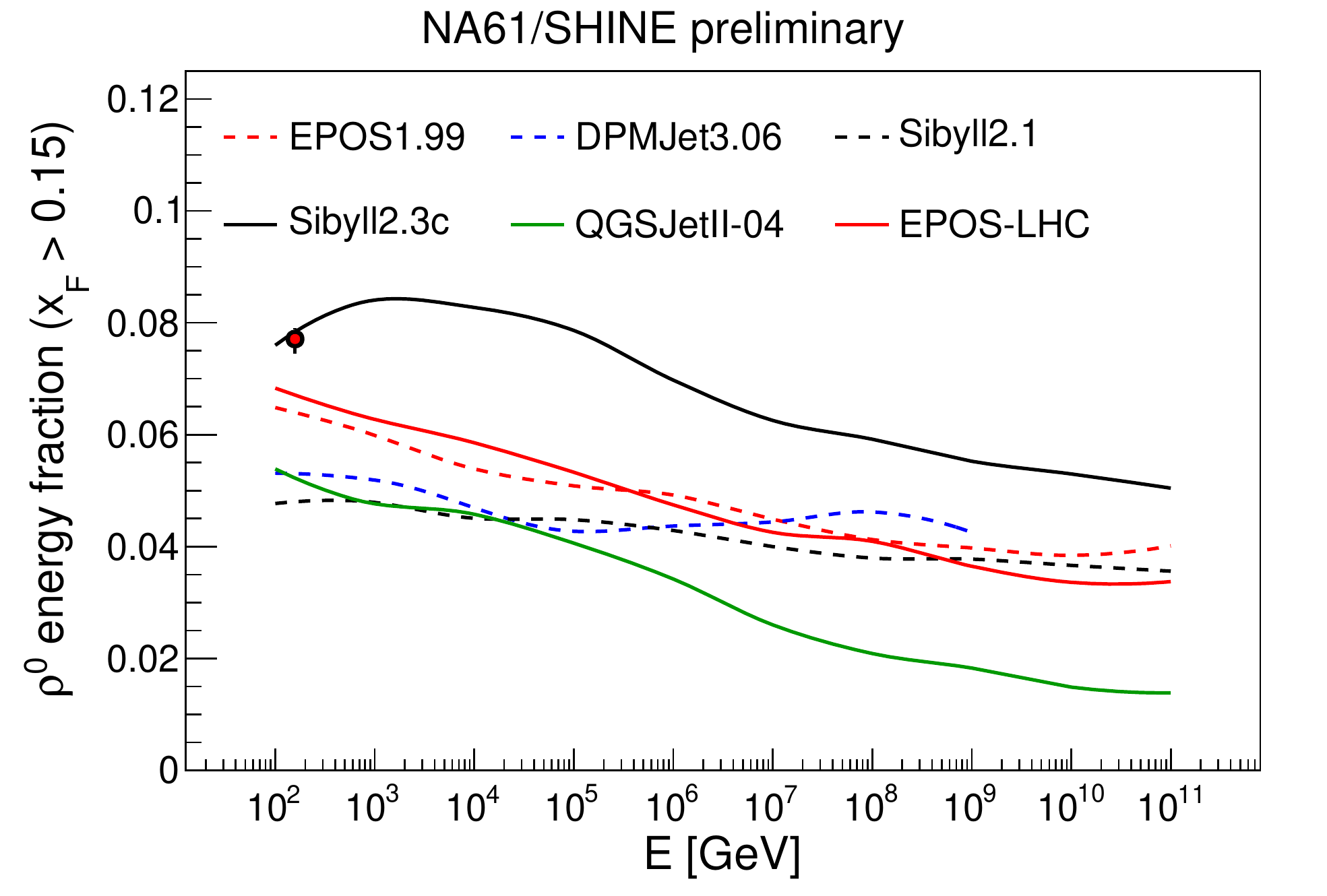}
 \caption{Energy fraction transferred to anti-protons (left) and $\rho^0$-mesons (right) as
   measured by \NASixtyOne (data points) and as predicted by hadronic interaction models
   over the whole range of beam energies relevant for air showers.}
  \label{fig:efrac}
\end{figure}

The total energy fractions transferred to anti-protons and $\rho^0$
mesons is displayed in Fig.~\ref{fig:efrac}. These fractions were
obtained by integrating the $p\text{d}n/\text{d}p$ spectra including
an extrapolation up to the full beam momentum~\cite{raulthesis, Prado:2017hub}. Note
that the anti-proton fraction measured here constrains the production
of p, \antiproton, n and $\bar{\text{n}}$ production. The sum of the
energy fractions of these particles is at about the same level as the
one going into $\rho^0$ mesons. The comparison to predictions of
hadronic models reveals that none of the existing attempts to describe
interactions in air showers succeeds to reproduce both energy
fractions at the same time. And although the energy dependence of the
energy fractions will remain a source of uncertainty for the
foreseeable future, a re-tune of these models to match the low energy
measurements from \NASixtyOne will significantly reduce the
uncertainties in predictions of muons in air showers.

\section{Results from Pilot Run on Nuclear Fragmentation}

Recent progress in the detection of Galactic cosmic rays with
space-based detectors such as PAMELA and AMS has lead to a wealth of
new data on the fluxes of leptons, nuclei and antiprotons from \GeV to
\TeV with an unprecedented percent-level precision. These new data
sets provide a unique diagnostic of cosmic-ray propagation in the
Galaxy and an opportunity to find signatures of astrophysical dark
matter annihilation. The amount of particle production in the Galaxy
depends on the integrated traversed matter density which can be
inferred from the ratio of secondary to primary cosmic rays such as
the ratio of Boron to Carbon if the nuclear fragmentation cross
section of the primary particles are known from laboratory
measurements. The current uncertainties in the modeling of the
propagation of cosmic rays in the Galaxy due to uncertainties of these
cross sections is however at the level of 10--20\%
\cite{2015A&A...580A...9G,Tomassetti:2017hbe,Reinert:2017aga,Evoli:2019wwu}
and thus much larger than the uncertainty of the cosmic-ray data itself.

To remedy this situation, the \NASixtyOne Collaboration is studying
the possibility of performing new precise measurements of nuclear
fragmentation at the CERN SPS. A pilot run~\cite{spscAddendumFrag}
took place in December 2018 and we took 3 days of test data for the
measurement of Boron production in C+p interactions.

\begin{figure}[t]
  \includegraphics[clip, rviewport = 0 0 1.02 1, width=0.5\linewidth]{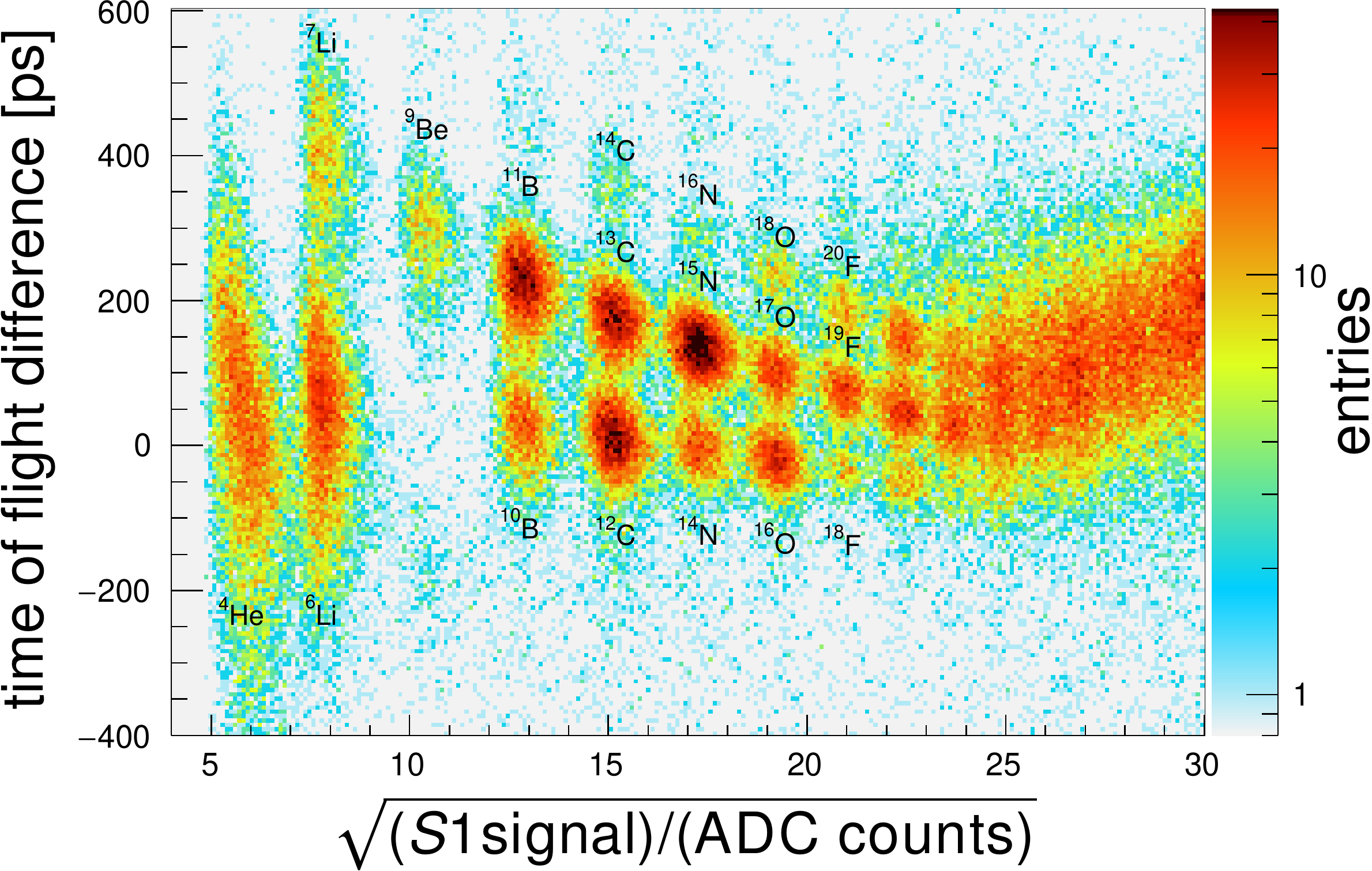}~\includegraphics[clip, rviewport = -0.02 0 1 1,width=0.5\linewidth]{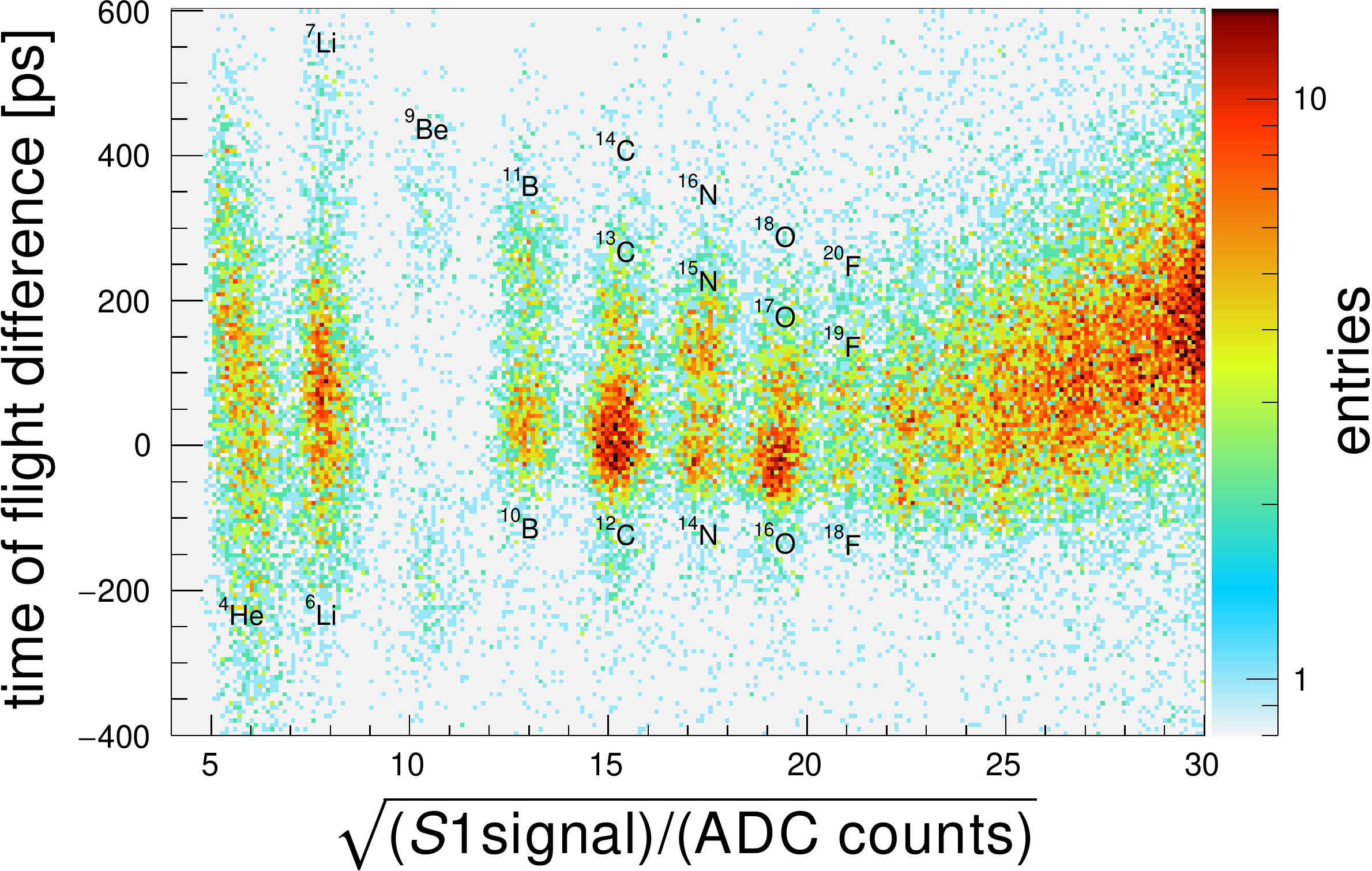}
  \caption{Beam composition recorded with free-running trigger
    detectors for different settings of the H2 beamline. Left:
    $E/Z=28$~GV, 20 cm primary Be target, right: $E/Z=27$~GV, 16 cm
    primary Be target. The signal on the x-axis is proportional to the
    charge of the beam particle, whereas the y-axis separates the
    isotopes masses via a measurement of the particle velocity.}
  \label{fig:compo}
\end{figure}

For this purpose, the SPS delivered secondary ions to our experiment
that were created by fragmenting a primary lead beam on a primary
beryllium target (see \cite{Strobele:2012zz} for a description of
previous data taking of NA61 with a secondary ion beam). A measurement
of the resulting beam composition at the experiment is shown in
Fig.~\ref{fig:compo} for two different beam rigidities $R=E/Z$ and
primary target thicknesses. As can be seen, even for the uncalibrated
online data shown here, the beam isotopes are well separated by
measuring their time of flight with two scinitillators separated by
236~m and the amplitude of the scintillator signals which is proportional
to the squared charge of the particle. The settings on the right panel
were used for the fragmentation run presented here as it maximized the
number of $^{12}$C ions in the beamline. The main beam trigger was
set to select carbon isotopes.

\begin{figure}[t]
  \includegraphics[clip, rviewport = 0 -0.25 1.1 1.05, width=0.5\linewidth]{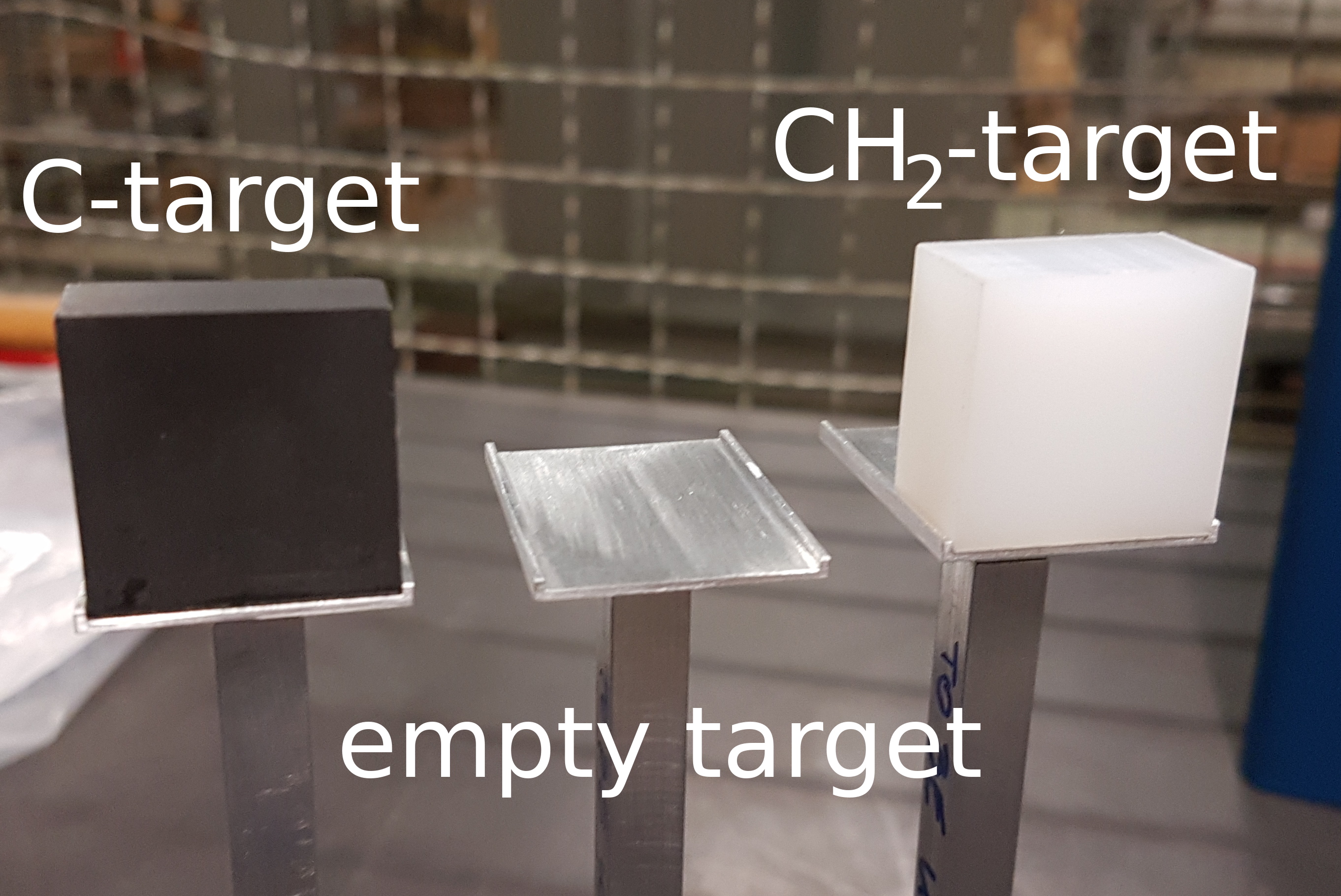}
  \includegraphics[width=0.5\linewidth]{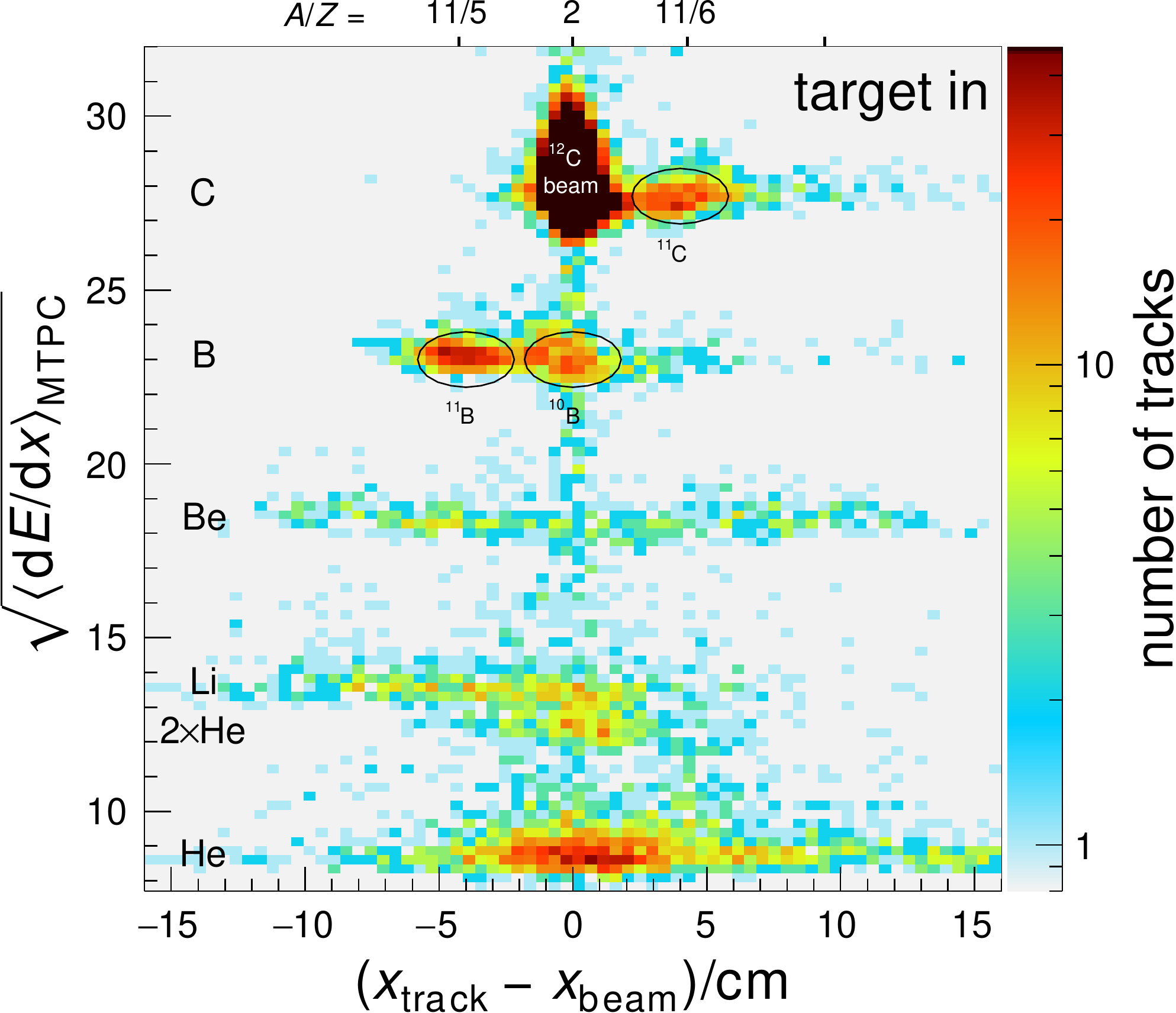}
  \caption{Left: target holder used for the data taking, right: detection of fragments in
    the TPC downstream of the magnetic field (see text)}
  \label{fig:targetmtpc}
\end{figure}

More than $1.1\times 10^6$ carbon beam triggers were recorded during
  three days of data taking alternating between three target positions
  as indicated on the left panel of Fig.~\ref{fig:targetmtpc}: A 1\,cm
  C-target, a 1.5\,cm polyethylene (CH$_2$) target and an empty target
  holder. The latter position is used to correct the measurement for
  out-of-target interactions and the subtraction of C+C interactions
  from C+CH$_2$ interactions yields the desired fragmentation cross
  sections in C+p interactions (details of the analysis can be found
  in~\cite{franziskathesis}).

  The identification of fragments downstream of the target is
  illustrated in the right panel of Fig.~\ref{fig:targetmtpc}. Here
  the x-axis shows the deviation of the measured track from the
  nominal beam position after the passage through the two
  superconducting magnets. This deviation is a measure of the rigidity
  of the particle relative to the beam rigidity and thus of the mass
  of the produced isotope. On the y-axis, the square root of the energy
  deposit in the TPC can be seen, which gives a measure of the charge
  of the particle. Since the data were taken in zero-bias mode, most
  tracks are in the $^{12}$C-peak appearing as a black area in the
  truncated color scale. $^{11}$C, $^{11}$B and $^{10}$B fragments are
  clearly visible and also Be and Li fragments, though the statistics
  of the pilot run is too small to discern the different isotopes of
  the latter by eye.

\begin{figure}[b]
  \centering
  \includegraphics[width=\linewidth]{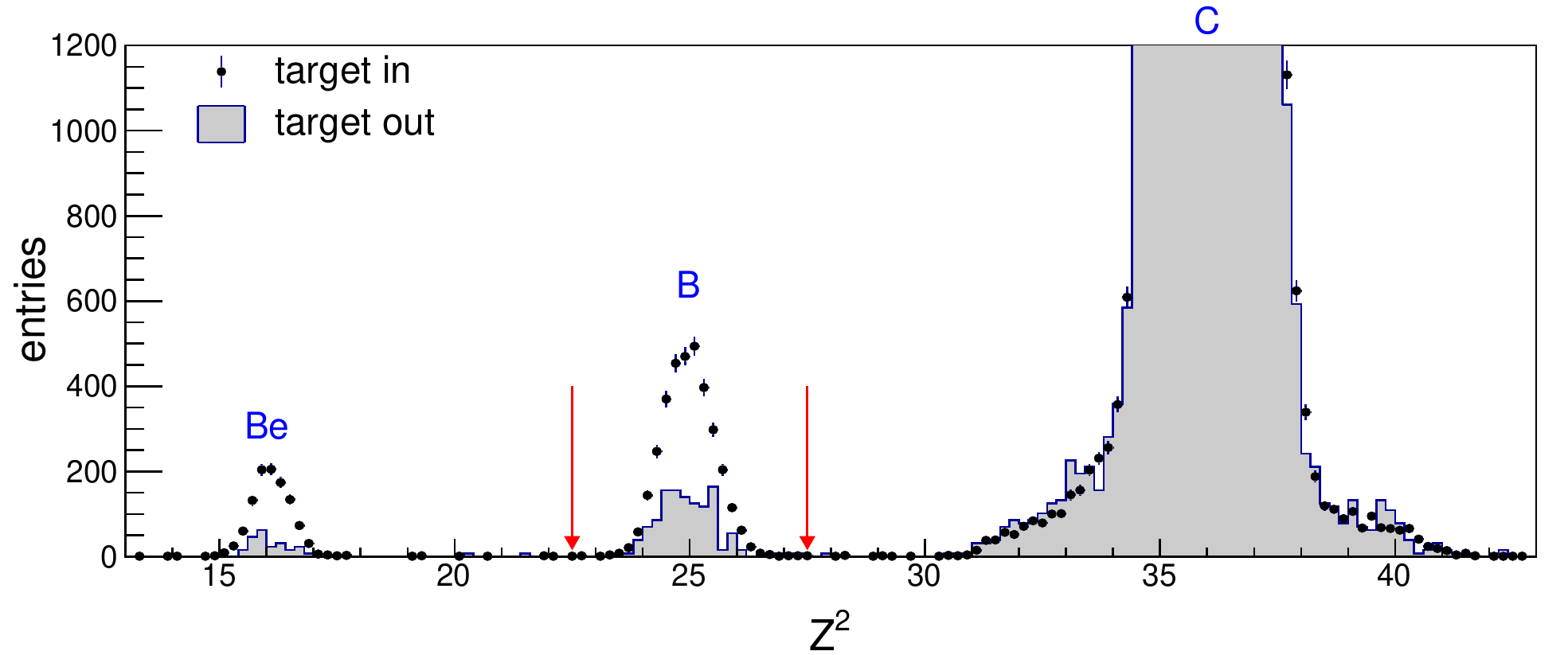}
  \caption{Distribution of the squared track charge measured with the TPC located
    about 10~m downstream of the target (see text).}
  \label{fig:mtpcsig}
\end{figure}

  As a first step, we concentrate the analysis on the sum of B
  isotopes.  The fully calibrated $Z^2$ distribution derived from the
  energy deposit in the TPC located 10~m downstream of the target is
  shown in Fig.~\ref{fig:mtpcsig}.  As can be seen, the different
  charges are well separated with almost no overlap between the
  elements. The gray histogram shows the distribution of charges
  without inserted target, i.e.\ isotopes that got produced in
  interactions with the detector material. The data points correspond
  to the charge distribution with inserted target and the red arrows
  indicate the range in $Z^2$ used to select B nuclei.

\begin{figure}[t]
  \centering
  \includegraphics[width=0.8\linewidth]{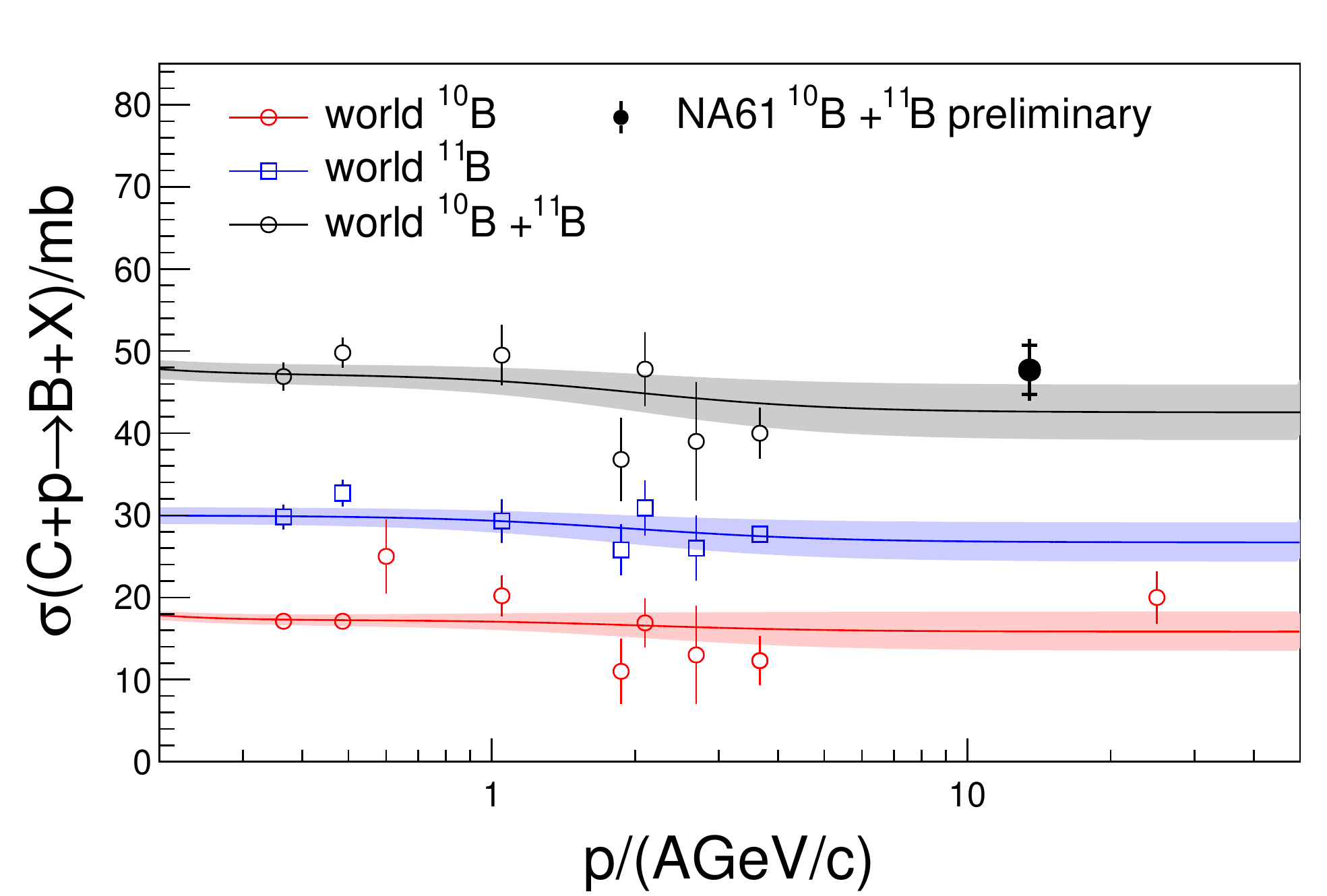}
  \caption[Result]{Measured fragmentation cross section of
    $\text{C}+\text{p} \rightarrow \text{B} +X$
    compared to previous measurements (open data points) and the
    parameterization of the fragmentation cross section from
    (see~\cite{Evoli:2019wwu} and references therein).}
  \label{fig:result}
\end{figure}

  From the number of B nuclei produced in C+C and C+CH$_2$
  interactions we derived a preliminary measurement of the
  fragmentation cross section of C on p to B:
  \begin{equation*}
    \sigma_{\text{C}+\text{p}\rightarrow \text{B} +X} = \left(47.7 \pm 3\,(\text{stat.}) \pm 2.3\,(\text{syst.})\right)\,\text{mb}.
  \end{equation*}
This measurement is compared to previous data in Fig.~\ref{fig:result}
and we find good agreement of our data with the extrapolation of the
cross section derived in~\cite{Evoli:2019wwu} indicated by the shaded
bands. It is worthwhile noting that the asymptotic value of the cross
section above 10~\AGeVc is very important to interpret the B/C ratio
measured by AMS up to $\gtrsim$TeV~\cite{Aguilar:2016vqr}. In this
energy range only one measurement from~\cite{Fontes:1977qq} was available
prior to our pilot run.

\section{Summary and Outlook}
The \NASixtyOne facility at the CERN SPS provides unique opportunities
to measure hadronic processes relevant for cosmic ray physics at fixed
target energies. With the updated corrections to the anti-proton
spectra presented here we conclude our extensive studies of the
interactions of pion projectiles, which are the most numerous
particles in cosmic-ray induced air showers. We presented preliminary
results from a pilot study for the measurement of the fragmentation of
nuclei relevant for the propagation of cosmic rays in the Galaxy.  The
quality of the test data demonstrates that \NASixtyOne is well
suited for these measurements.

As detailed in~\cite{geno}, a whole
range of nuclear reactions needs to be measured to allow for a precise
measurement of light secondary nuclei in cosmic rays. The variety of
secondary fragments delivered by the H2 beamline
(cf.~Fig.~\ref{fig:compo}) is ideal to collect this data at the SPS
with a single beam setting in a short time. Such a measurement
campaign is proposed to take place when the LHC (and thus SPS) resume
operation after the current ``long
shutdown''~\cite{Aduszkiewicz:2309890}.  The envisaged data taking
will benefit from the planned upgrade of the read-out system of NA61
to increase the possible data taking rate by a factor of
ten, bringing it up to 1~kHz. At this readout speed a modest beam time
of about two weeks of data taking with secondary ions from the SPS
would lead to a comprehensive measurement of the most relevant cross
sections for the secondary production of the cosmic-ray isotopes of
Li, Be, B, C and N.

\section*{Acknowledgments}
We would like to thank the organizers for the opportunity to present
these results at ICRC19 and Carmello Evoli for providing the data
tables and fits shown in Fig.~\ref{fig:result}. Furthermore we would
like to thank the CERN BE and EN Departments, in particular Reyes
Fernandez and Nikolaos Charitonidis, for their outstanding support of
NA61/SHINE without which none of the measurements presented here would
have been possible.

\bibliographystyle{na61Utphys}
\bibliography{na61References}

\end{document}

%% file: na61Abbreviations.tex
\newcommand{\eV}{\ensuremath{\mbox{e\kern-0.1em V}}\xspace}
\newcommand{\GeV}{\ensuremath{\mbox{Ge\kern-0.1em V}}\xspace}
\newcommand{\TeV}{\ensuremath{\mbox{Te\kern-0.1em V}}\xspace}
\newcommand{\MeV}{\ensuremath{\mbox{Me\kern-0.1em V}}\xspace}
\newcommand{\GeVc}{\ensuremath{\mbox{Ge\kern-0.1em V}\kern-0.1em/\kern-0.05em c}\xspace}
\newcommand{\GeVcc}{\ensuremath{\mbox{Ge\kern-0.1em V}\kern-0.1em/\kern-0.05em c^2}\xspace}
\newcommand{\AGeV}{\ensuremath{A\,\mbox{Ge\kern-0.1em V}}\xspace}
\newcommand{\AGeVc}{\ensuremath{A\,\mbox{Ge\kern-0.1em V}\kern-0.1em/\kern-0.05em c}\xspace}
\newcommand{\MeVc}{\ensuremath{\mbox{Me\kern-0.1em V}\kern-0.1em/\kern-0.05em c}\xspace}
\newcommand{\MeVcc}{\ensuremath{\mbox{Me\kern-0.1em V}\kern-0.1em/\kern-0.05em c^2}\xspace}

\newcommand{\pT}{\ensuremath{p_\text{T}}\xspace}

\def \pions{$\pi^\pm$\xspace}
\def \kaons{K$^\pm$\xspace}

\def \antiproton{$\bar{\text{p}}$\xspace}
\def \protons{p($\bar{\text{p}}$)\xspace}

\def \lambs{$\Lambda(\bar{\Lambda})$\xspace}
\def \kzeros{K$_{S}^{0}$\xspace}

\def \kaonstar{K$^{*0}$\xspace}
\def \rhozero{$\rho^{0}$\xspace}

\newcommand{\NASixtyOne}{NA61/\-SHINE\xspace}  
\newcommand{\CernVM}{\textsc{Cern\-\kern-0.05emVM}\xspace}